# Hydrogen Segregation in Palladium and the Combined Effects of Temperature and Defects on Mechanical Properties

Hieu H. Pham [1*], A. Amine Benzerga [2,3] and Tahir Çağın [1,3**]

[1] Department of Chemical Engineering
[2] Department of Aerospace Engineering
[3] Materials Science and Engineering
Texas A&M University, College Station, Texas 77843-3122, USA

**Abstract:** Atomistic calculations were carried out to investigate the mechanical properties of Pd crystals as a combined function of structural defects, hydrogen concentration and high temperature. These factors are found to individually induce degradation in the mechanical strength of Pd in a monotonous manner. In addition, defects such as vacancies and grain boundaries could provide a driving force for hydrogen segregation, thus enhance the tendency for their trapping. The simulations show that hydrogen maintains the highest localization at grain boundaries at ambient temperatures. This finding correlates well with the experimental observation that hydrogen embrittlement is more frequently observed around room temperature. The strength-limiting mechanism of mechanical failures induced by hydrogen is also discussed, which supports the hydrogen-enhanced localized plasticity theorem.

(*) Current address: Lawrence Berkeley National Laboratory, Berkeley, CA 94720 (hhpham@lbl.gov)

(**) Corresponding author: T.C. (tcagin@tamu.edu)

I.      INTRODUCTION

High activities of hydrogen in palladium have provided practical implications in many areas of materials science and chemistry, including catalysis [1-4], composite membranes [5-8], nanosensors [9-11], hydrogen storage and fuel cells [12-14]. In fact, a unique solubility of hydrogen in palladium was first reported as early as the mid-nineteenth century [15, 16]. However, it is believed that undesirable hydrogen absorption in crystals could substantially degrade their mechanical endurance, thus challenges the stability and performance of devices [17, 18]. Therefore, a fundamental understanding of hydrogen effect is necessary to help improve the materials' design and production.

It was accepted that elastic response of hydrogenated metals and alloys changes due to the lattice relaxation and local hydrogen excitation [19-22]. Several major hydrogen embrittlement mechanisms have been established: (i) hydrogen-enhanced localized plasticity [23]; (ii) stress-induced hydride formation [24]; and (iii) lattice, grain boundary (GB) and interface decohesion [25]. In general, diffusion of hydrogen atoms in crystals is enabled by their occupations of interstitial sites until they are trapped in open volumes such as vacancies and GBs. Additionally, GBs could also act as the transport channel of impurity species. In fact, presence of vacancies and GBs in metals and alloys could substantially alter their mechanical response, with or without the localization of impurity elements [26-28]. However, various studies have reported that hydrogen propagates GB cracks and reduces the ductility of metals and alloys [29-41]. It was suggested that hydrogen causes embrittlement by promoting the formation of microvoids and decreasing the cohesive strength of GBs.

Lately, atomistic simulations became convenient techniques for investigation of electronic structure and geometry of nanomaterials. In particular, Daw and Baskes developed the embedded-atom method (EAM), which was shown to successfully describe the properties of metal and alloy systems [42]. For instance, the method has been used by Zhong et al [43] to study the hydrogen-loaded palladium under tensile stress and they reported about the pre-melting disorder caused by hydrogen. In addition, first-principles calculations based on density functional theory have also been applied to examine the electronic origin of hydrogen-vacancy and hydrogen-GB complexes in several metal and alloys [44-47]. In this work, we use the classical molecular dynamics (MD) with embedded-atom method to direct theoretical investigations on hydrogen segregation in Pd using different crystal models (Table I). We have reported for the first time the hydrogen effect on Pd mechanical strength in a combined function with temperature and structural defects (vacancy and grain boundary). Throughout this work, low hydrogenated $PdH_x$ materials, with $x_H$ = [H]/[Pd] up to 0.1 is considered and the wide range of temperatures is covered (between 100 and 900 K).

## II. COMPUTATIONAL METHODS

We used the molecular dynamics method to compute equilibrium and mechanical properties of the H-Pd system. The nuclear motion of particles in MD simulation technique is based on classical mechanics [48], in which the interactions between atoms are characterized by a force field. In this work, we used the interaction potential from the EAM, which was developed by Daw and Baskes [42, 49]. This semi-empirical, many-body model describes the total energy of a metal atom as the energy obtained by embedding that atom into the local electron density of its atomic neighbors. Unlike pair potential models, in which the energy is just the sum over pair bonds, the total energy in EAM consists of an embedding energy term and the electrostatic interaction term.

$$E_c = \sum_i G_i \left( \sum_{j \neq i} \rho_j^a (R_{ij}) \right) + \frac{1}{2} \sum_{j \neq i} U_{ij} (R_{ij}) \qquad (1)$$

where the first term stands for the embedding energy and ρ is the background atomic electron density. The second term is related to the contribution from all cross-pair atomic interactions.

The EAM approach was reportedly capable of solving many problems of interests in metals and metal alloys, including defects, impurities, surface, fracture, etc. [50]. More specifically, it was successfully applied to metal hydride problems and hydrogen embrittlement phenomena [51-54]. We used in this work the EAM hydrogen-palladium interatomic potential developed by Zhou et al. [55]

A supercell model of 10,000 atoms was used to simulate the face-centered cubic (FCC) palladium (single crystal). In case of a bicrystal (i.e. FCC Pd bulk containing a grain boundary), we studied specifically a $\sum 5(210)$ grain boundary that separates the supercell into two differently-oriented grains of 5000 Pd atoms on each (Figure 1). This supercell size (10,000 Pd atoms) yields the convergence of GB surface energy as low as 0.1 meV/A$^2$. As a high-angle grain boundary, this GB plane offers considerable free volume for impurity transports and segregations and therefore could be potentially vulnerable to impurity-induced embrittlement.

Initial H-Pd configurations were prepared by randomly inserting H into interstitial or empty space inside Pd cells (single crystal, vacancy-containing and bicrystal supercells). Afterwards, the modulus and strength were studied by means of tensile test simulations. The MD time-step is chosen as 0.5 fs and the periodic boundary conditions are imposed in all x, y and z directions. At each temperature T and hydrogen content $x_H$, the reference zero-strain state was firstly obtained from constant temperature and stress ensembles by optimizing the volume of the supercell. The uniaxial tension is then applied in the z-direction (which is [210] for all supercell models, and normal to the GB plane in case of the bicrystal), and is increased gradually in steps of 50 MPa, followed by adequate relaxation to optimum volume (at that external pressure). The tensile strength $\sigma_{TS}$ is defined as the maximum tensile stress at which the system is sustained as stable. The tensile modulus in the applied tension direction is calculated by fitting the linear stress-strain relation [56] within a small range of strain (up to 0.5%):

$$E_z = \frac{d\sigma_z}{d\varepsilon_z} \quad (2)$$

where $\sigma_z$ and $\varepsilon_z$ correspond to stress and resulting strain along applied tension direction (z-axis).

Figure 2 shows the stress-strain relationship for a bicrystal at $x_H=0.1$ and T=300K, from which tensile modulus and tensile strength can be derived. When the applied stress is increased in the z-direction, boundaries of simulation box are allowed to shrink or dilate in order to maintain zero stress in x and y dimensions. In other words, in this method, we control the stress and calculate the subsequent strains (positive along z and negative along x and y), induced by that uniaxial stress. Having knowledge of these domain changes, we were also able to conduct a separate simulation, in which the strains are totally controlled and then stress could be obtained based on kinetic energy and interatomic interactions, using the canonical ensemble (NVT). This simulation approach (strain control) is actually a more precise reproduction of the tensile test experiment. However, the stress-strain curves obtained by theses two approaches (stress control and strain control) are virtually similar to each other (Figure 2). In the later method, the stress drops when the material is pulled up to a certain stretch, which indicates the breaking of some interatomic bonds. The ultimate tensile strength, or the maximum tensile stress at which the material can sustain stable deformation, is identical in both approaches. In this work, we adopted the stress-control procedure.

### III. RESULTS AND DISCUSSIONS

#### A. Hydrogen segregation at grain boundary

In the single crystal, the equilibrium distribution of hydrogen is expectedly uniform throughout the supercell. However, at the vicinity of grain boundary, the hydrogen segregation could be more complicated. At low temperature, hydrogen atoms have low diffusion coefficient. Upon temperature elevation, the diffusion would become active after H gains sufficient kinetic energy. Our simulation showed that GB might provide a great gradient for hydrogen absorption.

At high temperatures (above 400K), the equilibrium distribution of hydrogen is reached quickly in the bicrystal and tension also facilitates the hydrogen segregation at grain boundary. However, we found that, around room temperature, the saturation concentration of H at GB is relatively high. In addition, the H diffusion is quite active and far from the steady state, even after the equilibrium volume has been reached. Therefore, in order to obtain final H-Pd bicrystal configurations, we firstly heat the initial H-Pd systems up to 300K and maintain at this condition until the saturation of hydrogen at grain boundary was fully reached (approximately after lengthy $10^7$ time-steps). The structures used for tensile test then will be obtained from this configuration by heating (or cooling) and then sufficiently equilibrating at desired temperatures.

Figure 3a shows the final hydrogen distribution along z dimension at different temperatures (for the case of $x_H$= 0.06). Hydrogen distribution is monitored by dividing simulation box into 10 layers along z-direction, in which layers number one and six contain the GBs (showing high peaks in H content). We saw that between 100 and 300K, two GB-layers are able to trap totally up to 90% of total hydrogen in the bicrystal cell. Interestingly, H concentration at GBs reduces significantly after transition from 300 to 400K. The H contents at GB layers are also displayed in Figure 3b. H distribution is high and consistent in 100-300K and decreases above 300K (visualization of H distribution with $x_H$=0.06 is presented in Figure 4). This phenomenon can be explained that, at elevated temperature H gains sufficiently high kinetic energy, therefore is capable of escaping the H-GB trap. With increasing $x_H$, the real number of H atoms segregating at GB increases, but its relative portion compared to total amount of H would be highest at moderate $x_H$ = 0.04. In fact, there is always a certain amount of H left in crystal bulk and this amount doesn't increase when $x_H$ is low, because extra H will expectedly diffuse into grain boundary domains. When there is an excessive amount of hydrogen, the trapping energy of H at GB decreases and becomes more comparable to that at interstitial sites. For that reason, bulk H concentration will start to rise afterwards. In addition, our simulation also indicated that external tensile stress facilitates the hydrogen segregation at grain boundary.

When hydrogen diffusion is in active progress and H-GB accumulation increases, the simulation box also expands as the result of GB expansion. At that moment, it requires less effort to strain the system (in normal direction of GB). Therefore, the tensile test simulation should be conducted in the bicrystal cell only after full occupation of H at GB has been reached and H distribution is quite at its final stop.

## B. Further examination on the role of temperature, hydrogen content and structural defects

In this subsection, we will address the combined effects of temperature, hydrogen absorption and structural defects (vacancy and grain boundary) on mechanical properties of Pd crystals. It is believed that dopants and defects could substantially impact the properties of materials, from their electronic nature to macroscopic behavior. [27, 57, 58] In addition, open domain such as grain boundary and vacancy could be accessible to impurity absorption, as a result may lead to stress corrosion. In this work, for the particular investigation of mechanical properties, the tensile modulus and tensile strength were calculated. As the tensile stress is applied in the [210] direction, the quantities obtained will correspond to those in [210], i.e. tensile modulus $E_{[210]}$ and tensile strength $\sigma_{[210]}$ (hereafter shorten as E and σ).

In Table II and Table III we present the tensile modulus $E_{[210]}$ and tensile strength $\sigma_{[210]}$ at different T and $x_H$ for the single crystal; and the data is visualized in Figure 5. The maps reveal a monotonous decrease for both E and σ with respect to elevated T, which agrees with common trends that metallic materials soften at high temperature or reduced pressure [59, 60]. Also, the same behavior was observed for the increase in hydrogen concentration. The contour lines of σ and E surfaces on x-y plane are relatively straight, which suggests that variations of σ and E can be locally treated as plane surfaces with respect to T and $x_H$. However, on a bigger domain, those surfaces have some curvature, as the value of $\frac{\partial \sigma}{\partial T}$ and $\frac{\partial E}{\partial T}$ are not truly constant. While $\frac{\partial \sigma}{\partial T}$ decreases (in absolute value), the magnitude of $\frac{\partial E}{\partial T}$ increase with respect to higher $x_H$. In other words, the effects from temperature and hydrogen are not fully independent from each other.

Our calculations show that the presence of vacancy or grain boundary reduces both tensile modulus and tensile strength of the metal. For instance, at 300K and H-free state, tensile moduli $E_{[210]}$ of the vacancy-containing crystal and bicrystal are 82.02 and 73.58, respectively, compared to 87.02 GPa of the single crystal. The reductions in tensile strength are from 4.3 GPa (single crystal) to 3.45 (vacancy-containing crystal) and 3.75 GPa (bicrystal). Interestingly, we found that this trend (of decreasing modulus) is not preserved for the $\sum 5(100)$ Pd GB. In fact, Zugic and coworkers reported an increase of Young modulus across a $\sum 5(100)$ grain boundary in nickel from their experiments [56]. This discrepancy between $\sum 5(100)$ and $\sum 5(210)$ could be explained as the geometry factor of each GB type, in which a portion of atoms situated at $\sum 5(100)$ GB in FCC crystals has short interatomic distance and overweighed the atoms with expanded distances in their contribution to elastic response, due to the anharmonicity of interatomic potential [56]. Meanwhile, the large grain mismatch and big free volume in the $\sum 5(210)$ GB, instead, can result in material softening.

In Figure 6, we present the tensile modulus and tensile strength of the vacancy-containing crystal and bicrystal supercells, for hydrogen content $0 \leq x_H \leq 0.1$ and temperatures from 100 to 900K. As discussed in the precious subsection, hydrogen atoms are distributed quite uniformly in single crystals (including ones with vacancy). The hydrogen concentration in those cases can be referred as a global quantity. Contrariwise, as hydrogen accumulates densely at GB, the local concentration of H at GB in the bicrystal cell indeed far exceeds its global value.

In the presence of defects (a vacancy or a $\sum 5(210)$ GB), $\sigma$ and E decrease monotonously with both increasing T and $x_H$, similar to what was seen previously for the single crystal. These fairly linear behaviors of $\frac{\partial \sigma}{\partial T}$ and $\frac{\partial E}{\partial T}$ (with respect to $x_H$) imply nearly constant second derivatives $\frac{\partial^2 \sigma}{\partial C \partial T}$ and $\frac{\partial^2 E}{\partial C \partial T}$ and this suggests one to fit the data using $\sigma(T, x) = \sigma_1 T + \sigma_2 x + \sigma_3 Tx + \sigma_4$ and $E(T, x) = e_1 T + e_2 x + e_3 T + e_4$ for tensile strength and tensile modulus, respectively, as functions of temperature T and hydrogen content $x_H$. These fitting coefficients were tabulated in Table IV.

The values of $\sigma_4$ and $e_4$ correspond to $\sigma$ and E at infinitesimally low temperature and hydrogen-free state, which stand for characterization of the crystal defects only. These values drop, for instance, by 16% in tensile modulus and 11% in tensile strength with the effect of grain boundary. Coefficients $\sigma_1$, $\sigma_2$, $e_1$, $e_2$ come from the independent effects of temperature T and hydrogen $x_H$ and their negative values indicate that the increasing T and H absorption degrade the mechanical properties. The weight of temperature effect ($\sigma_1$) to tensile modulus doesn't change regardless of structural characterization. Also, with the introduction of defects, the hydrogen effect on tensile modulus decreases but its effect on tensile strength varies just slightly. Coefficients $\sigma_3$ and $e_3$ are added as correlated terms that characterize the contribution from T or $x_H$ when the other one is changing. Positive $\sigma_3$ also plays its role in describing a concave surface of $\sigma_{[2\,1\,0]}$, and the convex surface of $E_{[2\,1\,0]}$ is characterized by a negative $e_3$. Note that, these terms have no significant change for different structures.

To make a specific comparison, in Figure 7 we presented their stress-strain behaviors at 300K and $x_H = 0.1$, in which the curves for single crystal and vacancy crystal are quite compatible. The bicrystal has a lower tensile modulus (57.2 GPa), compared to the single (73.5 GPa) and vacancy ones (70.5 GPa). There is also a big degradation (34%) from that of the H-clean single crystal (87 GPa). Interestingly, the bicrystal has the same tensile strength with the other two, regardless of being softer. In other words, the bicrystal is more ductile, but still comes to unstable state at the same stress compared to the others, under high H absorption ($x_H = 0.1$ cause a 40% reduction in tensile strength). The explanation for a higher ductility in bicrystal is that the mechanical degradation in a single crystal occurs throughout the whole cell, as opposed to one specific location (GB in the bicrystal cell).

In Figure 8 we display the atomic configuration of defective Pd atoms in different crystals, respectively, with $x_H = 0.1$ at zero tress (Figure 8a-c) and corresponding maximum tensile strength (Figure 8d-f). The concentration of "defective" Pd induced by vacancy and hydrogen binding is pretty high and this can be the reason why the vacancy-containing crystal has a lower tensile strength compared to the bicrystal and single crystal. We also observed the formation of dislocations (stacking fault) at grain boundary at high external tensile stress (Figure 8f). This specific atomic plane was detected to be (1 1 -1) and belongs to the general family of close-packed slip planes in FCC metals. This emission of dislocations could infer about hydrogen-enhanced localized plasticity (HELP) mechanism [23] and that the failures at stress above tensile strength are of plastic nature rather than brittle. Also, it aligns with several experimental observations that fracture process occurs in the vicinity of grain boundary by highly localized plasticity, as opposed to embrittlement along the grain boundary [23, 61].

## IV. CONCLUSIONS

Grain boundaries are easily exposed to hydrogen absorption due to high segregation energy, which could accommodate a significant portion of hydrogen in nanocrystalline materials. Even at ambient conditions, H atoms gain sufficient kinetic energy to diffuse from bulk phase into grain boundarie domains; subsequently, high H-GB concentration could be observed. However, the H segregation at GBs remarkably decreases as temperatures increase, due to active H diffusion and high kinetic energy. This result supports experimental observations that materials are more susceptible to hydrogen-induced failure at ambient temperatures [34].

In general, the tensile strength and tensile modulus decrease monotonously with increasing temperature and H content. As shown for the bicrystal, hydrogen segregation may induce plastic failure in the vicinity of the GB, as opposed to only causing embrittlement along the GB. In addition, the accumulation of H at GBs could be facilitated as a result of external tensile stress; therefore the combination of factors such as moderate heat and pressure in turn may stimulate the failure of polycrystalline materials at a much higher pace.

**Acknowledgement.** The research here was supported by ONR. The authors would also acknowledge the Texas A&M Supercomputing Facility (http://sc.tamu.edu/) for providing computing resources. We also would like to thank Dr. X. W. Zhou for providing the Pd-H interatomic potential that was used in this study.

TABLE I. Free volume and free surface in single crystal, vacancy-containing crystal and bicrystal supercells (at 300K)

| Crystal models | total vol., nm$^3$ | % free vol. | free surface, nm$^2$ |
|---|---|---|---|
| Single crystal ($Pd_{10000}$) | 148.28 | 1.37 | 23.79 |
| 1%-vacancy crystal ($Pd_{9900}Vac_{100}$) | 147.41 | 1.73 | 42.39 |
| Bicrystal ($Pd_{5000}Pd_{5000}$) | 148.95 | 1.63 | 36.71 |

TABLE II. Tensile modulus $E_{[2\,1\,0]}$ of Pd single crystal

| T | clean Pd | $PdH_{0.02}$ | $PdH_{0.04}$ | $PdH_{0.06}$ | $PdH_{0.08}$ | $PdH_{0.10}$ |
|---|---|---|---|---|---|---|
| 100 | 96.44 | 92.77 | 89.55 | 86.36 | 83.79 | 80.99 |
| 200 | 91.73 | 88.70 | 86.07 | 83.90 | 80.68 | 77.17 |
| 300 | 87.02 | 83.95 | 80.71 | 79.24 | 76.83 | 73.53 |
| 400 | 82.93 | 81.37 | 78.26 | 74.47 | 72.20 | 69.26 |
| 500 | 78.61 | 75.93 | 74.35 | 69.45 | 67.87 | 62.10 |
| 600 | 74.38 | 69.83 | 67.58 | 64.03 | 58.88 | 56.52 |
| 700 | 67.91 | 64.77 | 61.84 | 59.68 | 54.04 | 51.35 |
| 800 | 63.93 | 60.23 | 57.13 | 53.49 | 49.34 | 45.02 |
| 900 | 58.26 | 54.20 | 51.27 | 47.88 | 44.16 | 39.16 |
| $\frac{\partial E}{\partial T}$, $10^{-2}$ | -4.71 | -4.83 | -4.81 | -4.91 | -5.19 | -5.35 |

TABLE III. Tensile strength $\sigma_{[2\,1\,0]}$ in Pd single crystal

| T | clean Pd | $PdH_{0.02}$ | $PdH_{0.04}$ | $PdH_{0.06}$ | $PdH_{0.08}$ | $PdH_{0.10}$ |
|---|---|---|---|---|---|---|
| 100 | 5.30 | 4.80 | 4.30 | 4.15 | 3.65 | 3.45 |
| 200 | 4.70 | 4.30 | 4.05 | 3.70 | 3.30 | 2.99 |
| 300 | 4.30 | 3.90 | 3.50 | 3.10 | 3.00 | 2.65 |
| 400 | 3.90 | 3.45 | 3.15 | 2.80 | 2.45 | 2.45 |
| 500 | 3.55 | 3.25 | 2.95 | 2.55 | 2.20 | 2.10 |
| 600 | 3.25 | 2.95 | 2.65 | 2.35 | 2.10 | 2.00 |
| 700 | 2.90 | 2.70 | 2.45 | 2.15 | 1.95 | 1.80 |
| 800 | 2.65 | 2.45 | 2.20 | 2.00 | 1.80 | 1.65 |
| 900 | 2.30 | 2.20 | 1.95 | 1.70 | 1.65 | 1.50 |
| $\frac{\partial \sigma}{\partial T}$, $10^{-3}$ | -3.60 | -3.14 | -2.93 | -2.88 | -2.49 | -2.33 |

TABLE IV. Fitting coefficients to $\sigma(T, x) = \sigma_1 T + \sigma_2 x + \sigma_3 Tx + \sigma_4$ and $E(T, x) = e_1 T + e_2 x + e_3 Tx + e_4$

| Fitting coef., GPa | single crystal | 1%-vacancy crystal | bicrystal |
|---|---|---|---|
| $e_1$, $10^{-2}$ | -4.655 | -4.368 | -4.081 |
| $e_2$ | -1.291 | -1.191 | -1.295 |
| $e_3$, $10^{-2}$ | -0.062 | -0.058 | -0.122 |
| $e_4$ | 101.313 | 96.442 | 84.343 |
| $\sigma_1$, $10^{-2}$ | -0.349 | -0.337 | -0.345 |
| $\sigma_2$ | -0.20 | -0.132 | -0.140 |
| $\sigma_3$, $10^{-2}$ | 0.012 | 0.010 | 0.009 |
| $\sigma_4$ | 5.353 | 4.541 | 4.759 |

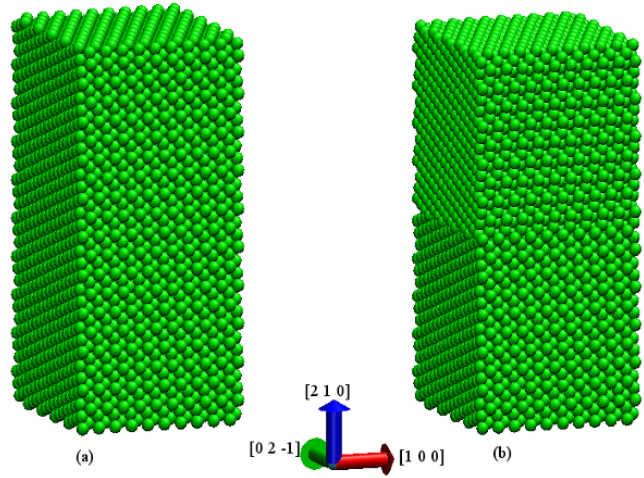

FIG. 1. (a) A 10000-atom model of a FCC $Pd_{10000}$ single crystal tensile specimen. (b) $Pd_{5000}Pd_{5000}$ bicrystal supercell with a $\sum 5(210)$ grain boundary.

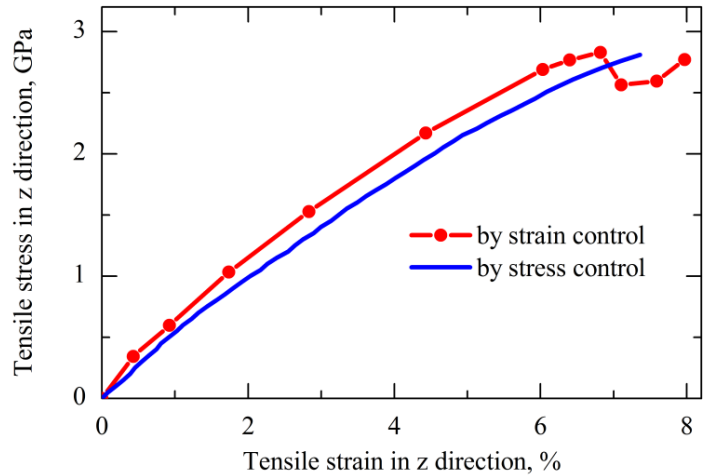

FIG. 2. Stress-strain curves for a Pd single crystal at T=300K and $x_H$=0.1, obtained using two simulation approaches. Under stress-control, the tensile stress is increased in z-direction and kept as zero in others. Under strain-control, a constant temperature and constant volume ensemble is used to calculate the stress.

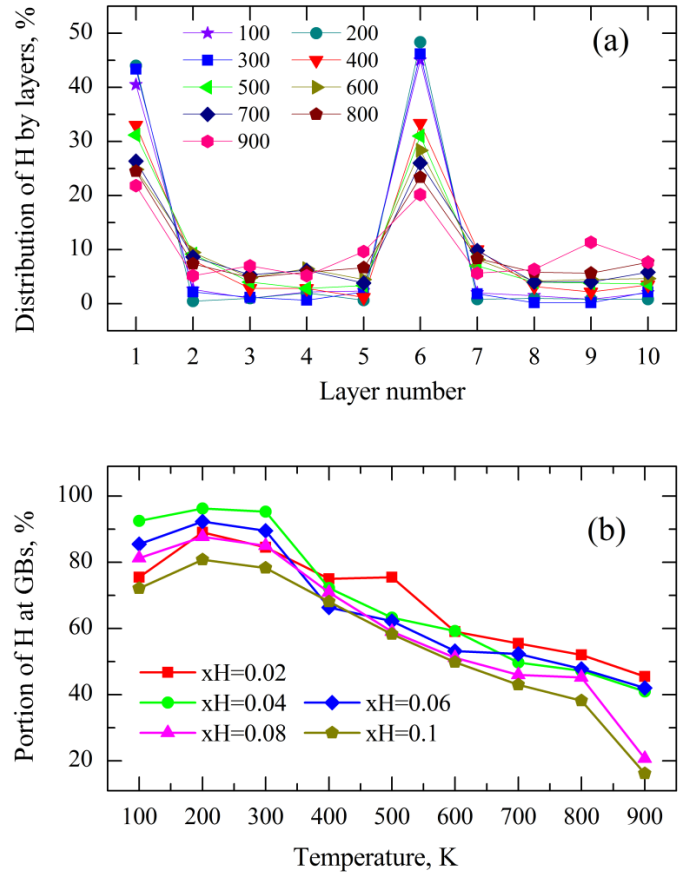

FIG. 3. Distribution of hydrogen inside the bicrystal supercell. (a) Distribution of H by layers when H concentration is 6% atomic. The whole supercell model is divided by 10 layers, in which layer number 1 and 6 contain the grain boundary. The total hydrogen content at GB layers is very high, especially at 300K and below. At 400K and higher, it decreases. (b) Comparison of H content at GB layers for different hydrogen concentration.

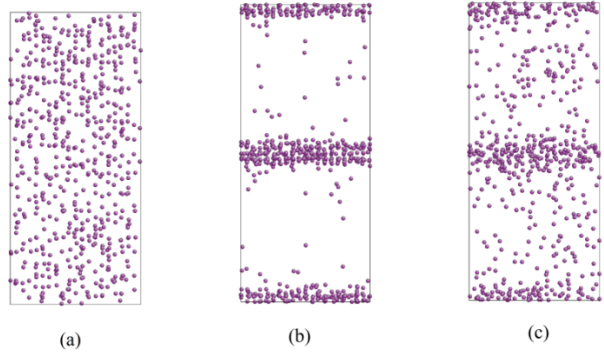

FIG. 4. Demonstration of hydrogen segregation at $x_H$=0.06 (Pd atoms are not shown). (a) Fairly uniform distribution of H in a single crystal at 300K. (b) 90% of H is trapped at the GBs in the bicrystal at 300K. (c) Distribution of H in the bicrystal supercell at 600K

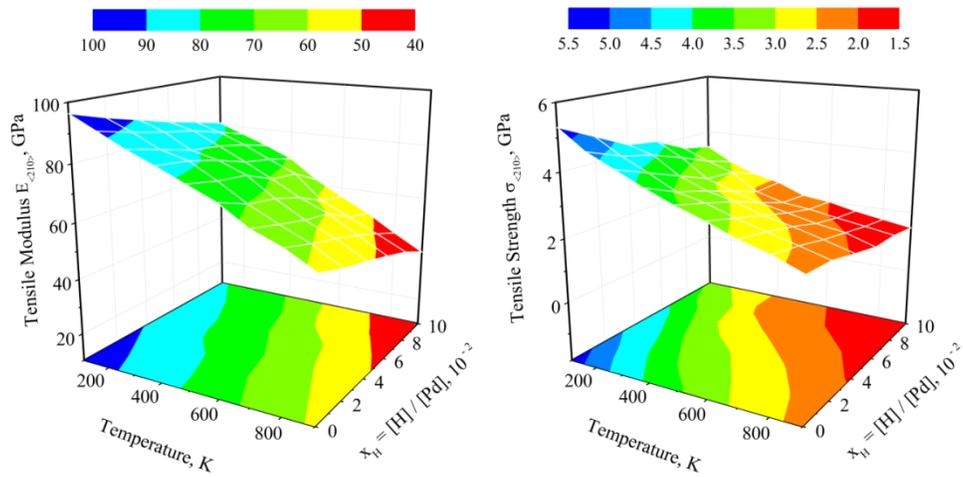

FIG. 5. Maps showing the linear variations of tensile modulus $E_{[210]}$ and tensile strength $\sigma_{[210]}$ as functions of temperature and H content $x_H$

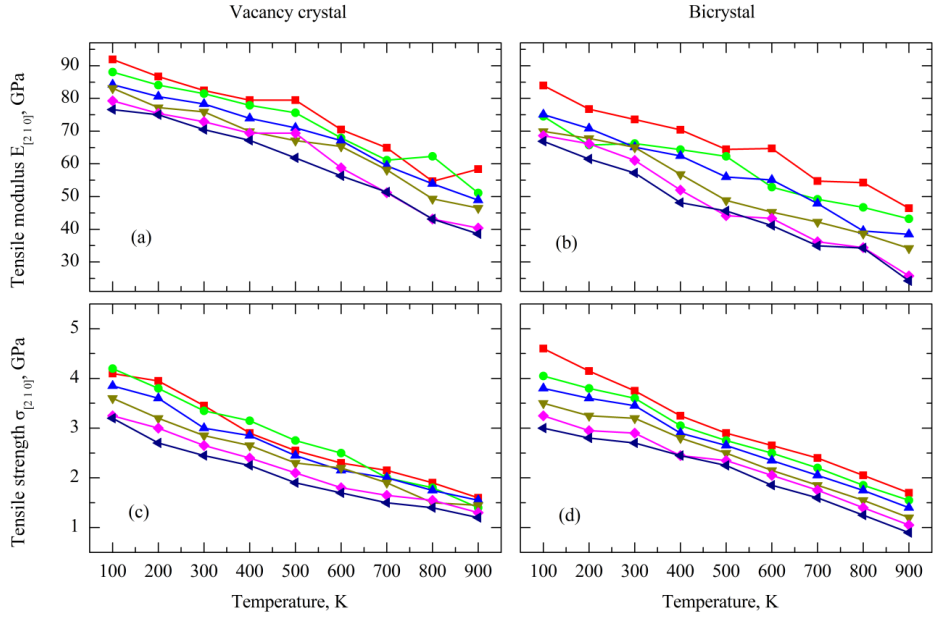

FIG. 6. Tensile modulus $E_{[2\,1\,0]}$ (a and b) and tensile strength $\sigma_{[2\,1\,0]}$ (c and d) as functions of T and total H content in high-vacancy crystal and bicrystal: ■ $x_H = 0$, ● $x_H = 0.02$, ▲ $x_H = 0.04$, ▼ $x_H = 0.06$, ♦ $x_H = 0.08$, ◄ $x_H = 0.1$

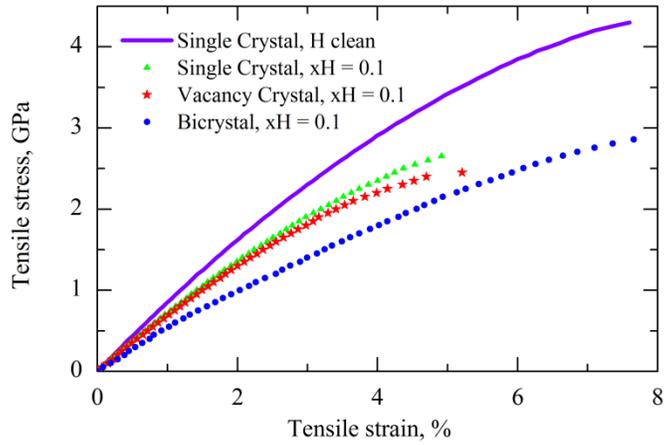

FIG. 7. Comparison of stress-strain curves for different H-Pd systems (T = 300K)

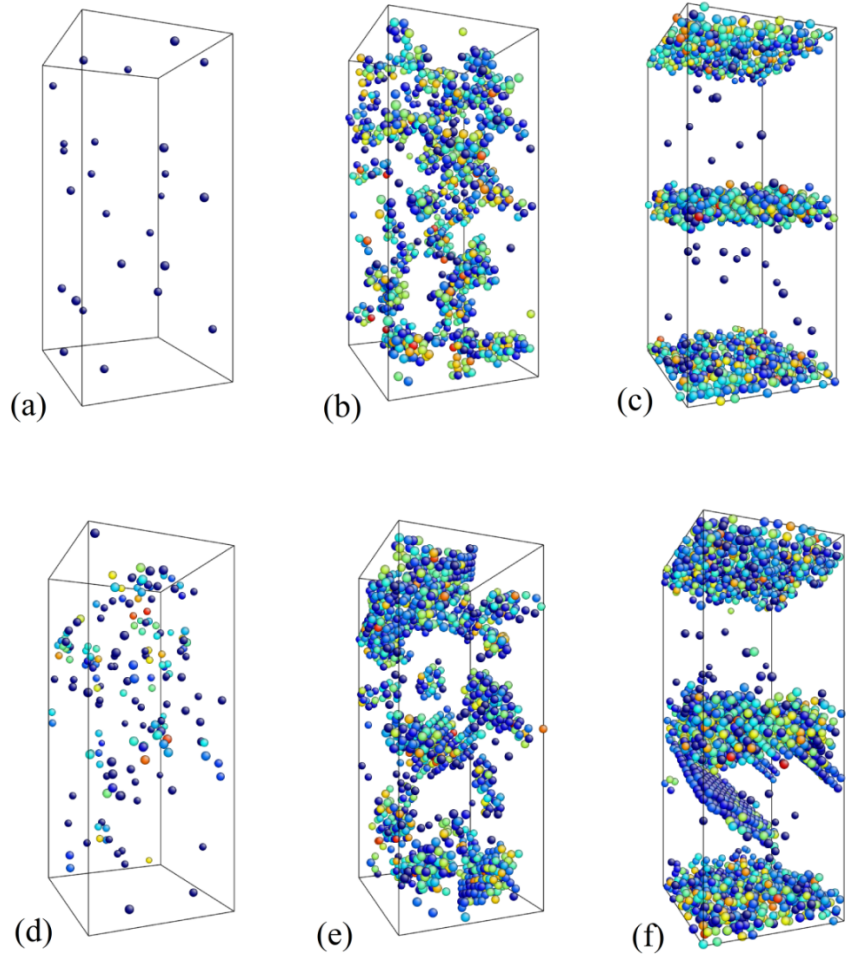

FIG. 8. Atomic configurations of Palladium with $x_H$=0.1 in single crystal, 1%-vacancy-containing crystal and bicrystal cells, respectively, at zero tensile stress (a,b,c) and ultimate tensile strength (d,e,f). Only defective Pd atoms are shown based on centrosymmetry parameters. The visualization uses AtomEye [62]